\documentclass{iau}
\usepackage{graphicx}
\usepackage{natbib}

\title[HD~82943] 
{On the orbital structure of the HD~82943 multi-planet system}

\author[R.V. Baluev \& C. Beaug\'e]   
{Roman V. Baluev$^{1,2}$ \and Cristian Beaug\'e$^3$}

\affiliation{$^1$Central Astronomical Observatory at Pulkovo of Russian Academy of Sciences,
Pulkovskoje shosse 65, St Petersburg 196140, Russia \\[\affilskip]
$^2$Sobolev Astronomical Institute, St Petersburg State University, Universitetskij
prospekt 28, Petrodvorets, St Petersburg 198504, Russia \\ email: {\tt r.baluev@spbu.ru} \\[\affilskip]
$^3$Instituto de Astronom\'ia Te\'orica y Experimental, Observatorio Astron\'omico,
Universidad Nacional de C\'ordoba, \\ Laprida 854, (X5000BGR) C\'ordoba, Argentina}

\pubyear{2014}
\volume{310}  
\pagerange{xxx--xxx}
\jname{Complex Planetary Systems}
\editors{Z. Knezevic \& A. Lema\^itre}
\begin{document}

\maketitle

\begin{abstract}
HD~82943 hosts a mysterious multi-planet system in the 2:1 mean-motion resonance
that puzzles astronomers for more than a decade. We describe our new analysis of all radial
velocity data currently available for this star, including both the most recent Keck
data and the older but more numerous CORALIE measurements.

Here we pay a major attention to the task of optimal scheduling of the
future observation of this system. Applying several optimality criteria,
we demonstrate that in the forthcoming observational season of HD~82943 (the
winter 2014/2015) rather promising time ranges can be found. Observations of the
near future may give rather remarkable improvement of the orbital fit, but only if
we choose their time carefully.
\keywords{stars: planetary systems - stars: individual: HD~82943 - techniques: radial velocity
- methods: data analysis - methods: statistical}
\end{abstract}

\firstsection 
\section{Introduction}
This paper can be treated as an addition to our recent work \citep{BaluevBeauge14} devoted
to a reanalysis of the radial velocity (RV) data for a unique multi-planet host star
HD~82943. Here we only briefly summarize the most important of our
previous results (Sect.~\ref{sec_oldres}), and present new ones, related
to seeking the optimal observation dates for this star (Sect.~\ref{sec_plan}).

\section{Main results of the RV data analysis}
\label{sec_oldres}
In our work we used the entire set of the RV data currently available for HD~82943
in the public literature. These include the old CORALIE \citep{Mayor04} and the recent Keck
\citep{Tan13} data. The Keck data were separated in two independent subsets that
were acquired before and after a hardware upgrade. The primary results
concerning our re-analysis of these data are described in \citep{BaluevBeauge14}. Thus
we do not duplicate this discussion here, except for a brief summary of the conclusions:
\begin{enumerate}
\item The Keck and CORALIE data are not in a good agreement with each
other: fitting the entire data set plainly leads to a severely
unstable orbital configuration of the two major planets \emph{b} and \emph{c}.

\item One of the reasons for this mutual inconsistency is the likely presence of
an additional systematic variation in the CORALIE (but not Keck) data with a period close to
a year. Likely, this variation appeared due to some imperfections of the
spectrum processing algorithm used for CORALIE.

\item After removal of the CORALIE annual variation, the RV data still contain a significant
periodicity with a period of $\sim 1000$~d. We interpret this periodicity as a hint of the
third planet that was previously suspected by \citet{GozdKon06} and \citet{Beauge08}.

\item An RV fit implying a stable planetary configuration can be obtained only by including
both the third planet and the CORALIE annual term in the RV model. Without these terms, the
nominal (best fitting) solution appears unstable due to an antialigned initial apsidal
state of the two major planets, and forcing this configuration to be stable would infer an
unsuitably large shift of the fit from its nominal position.

\item The planets in the best fitting configuration lie near the
three-planet resonance with the periods ratio $P_c:P_b:P_d\approx 1:2:5$. The motion of the
first two planets is truly resonant (a libration) in the vicinity of
an aligned Apsidal Corotation Resonance (ACR), while the orbit of the third planet
is rather uncertain and its orbital evolution can represent a libration (i.e. a true
resonance) as well as a circulation (not a genuine resonance). The structure
of the dynamical space in the vicinity of the third planet's orbit is pretty complicated
and involves resonant domains intervening with the non-resonant ones.
\end{enumerate}

\section{Optimal planning of the future RV observations}
\label{sec_plan}
Clearly, the orbital and dynamical structure of the HD~82943 system is still
rather uncertain. In view of this, it may be useful to apply some optimal planning
routines, in order to predict the time segments in future in which the new RV observations
would improve or knowledge about the system, as well as to identify the time
ranges where the new observations would be almost useless.

We solve this task by means of the optimal planning approaches described
in \citep{Baluev08a}. In this method we should select the entire set of the fitted
parameters, or any their subset, or even a set of some other quantities expressed by smooth
functions of the original parameters. Our goal is to find an optimal
time for a new observation in the future, in order to achieve a maximum reduction of the
uncertainties in the targeted quantities. Here we adopt the so-called D-optimality
criterion, in which the ``reduction'' of a multi-dimensional uncertainty is treated in
terms of the volumes ratio for the relevant uncertainty ellipsoids (or determinants of the
relevant covariance matrices).

\begin{table}
\begin{center}
\caption{Prescribed observations scheduling goals for HD~82943.}
{\scriptsize
\begin{tabular}{|l|c|c|}
\hline
scheduling goal   & critical parameters to refine        & their number \\
\hline
goal 1 & parameters of all three planets & $14$ \\
goal 2 & common orbital inclination           & $1$  \\
goal 3 & location relatively to the two-planet ACR (see text) & $4$  \\
goal 4 & parameters of the third planet  & $3$  \\
\hline
\end{tabular}
}
\label{tab_goals}
\end{center}
\end{table}

In this work we consider the three-planet fit with the eccentricity of the third
planet always fixed at zero. Otherwise this eccentricity is ill determined and generates
dramatic non-linearity effects, which are not desirable. Four sets of target quantities to
refine were considered in this work, defining four scheduling
``goals''. These goals are described in Table~\ref{tab_goals}. The goal~3 from
this table is defined in a rather complicated manner. Its purpose is to refine our
knowledge about the position of the dynamical system relatively to the ACR of the two major
planets. This information is important for the long-term dynamics and the stability of the
system \citep{Beauge03}. In this case the set of target quantities
includes $4$ partial derivatives of the averaged Hamiltonian of the two major planets. When
all these derivatives vanish, we deal with an exact ACR, so by reducing their uncertainties
we can improve our knowledge about whether the system is close to or far from the ACR. The
derivatives are computed using the method from \citep{Baluev08b}.

\begin{figure}[b]
\begin{center}
 \includegraphics[width=0.99\textwidth]{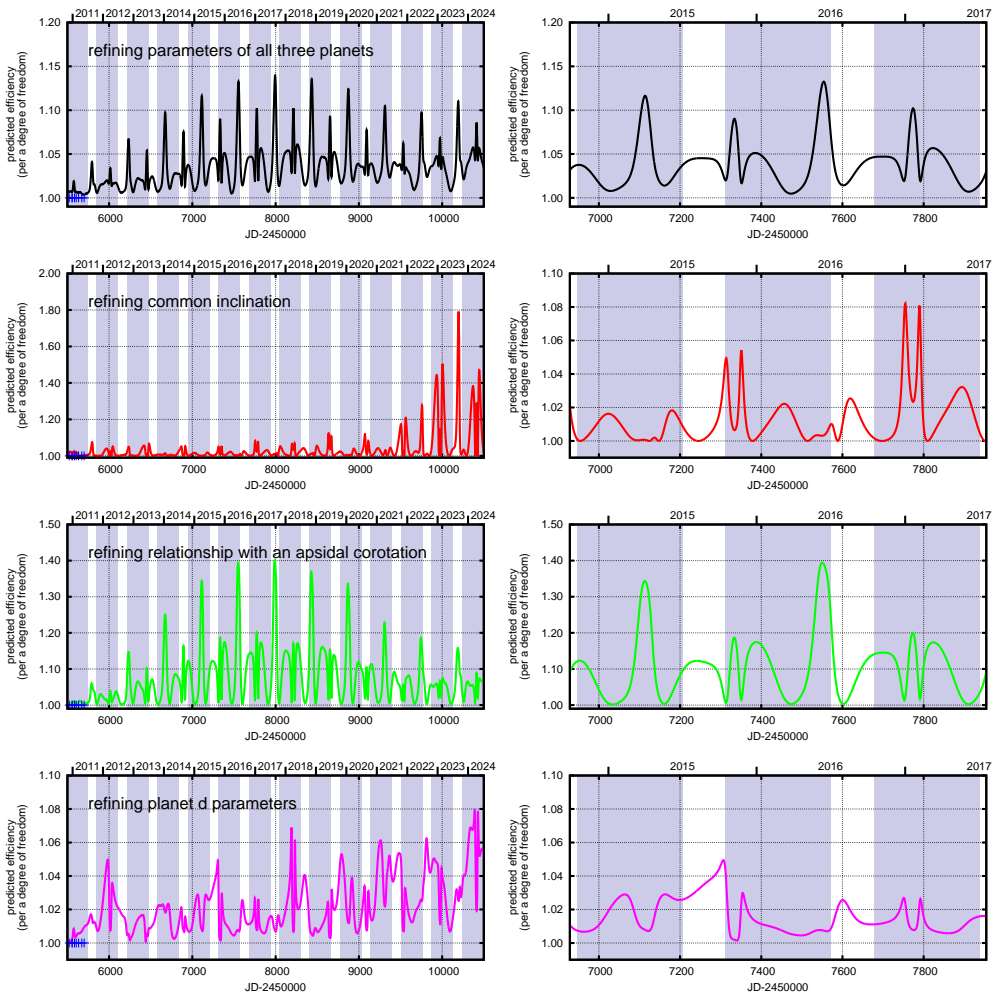} 
 \caption{The predicted Keck observations efficiency for scheduling goals from Table~\ref{tab_goals}.}
   \label{fig_sched}
\end{center}
\end{figure}

The results of the computation are shown in Fig.~\ref{fig_sched}, where we plot the graphs
of a function that indicates how much the uncertainty of the target quantities would reduce,
provided that we make a single observation at the time given in the abscissa. This relative
reduction is normalized so that it corresponds to a single degree of freedom in the set of
target quantities. The vertical fringes label the yearly seasons when the star can (darker
bands) or cannot (white bands) be actually observed.

\begin{table}
\begin{center}
\caption{Optimal observation dates for HD~82943.}
{\scriptsize
\begin{tabular}{|ll|ll|ll|ll|}
\hline
\multicolumn{2}{|c|}{scheduling goal 1} & \multicolumn{2}{|c|}{scheduling goal 2} & \multicolumn{2}{|c|}{scheduling goal 3} & \multicolumn{2}{|c|}{scheduling goal 4} \\
\hline
JD-$2450000$    & max. eff. & JD-$2450000$    & max. eff. & JD-$2450000$    & max. eff. & JD-$2450000$    & max. eff. \\
\hline
\multicolumn{8}{|c|}{observational season of 2014/2015}\\
begin -- $6988$ & $1.038$ & $7024 \pm 27$   & $1.016$ & begin -- $6985$ & $1.123$ & $7055 \pm 32$   & $1.029$ \\
$7113 \pm 15$   & $1.116$ & $7182 \pm 17$   & $1.018$ & $7113 \pm 17$   & $1.343$ & $7140$ -- end   & $1.029$ \\
$7193$ -- end   & $1.035$ & -               & -       & $7201$ -- end   & $1.095$ & -               & -       \\
\multicolumn{8}{|c|}{observational season of 2015/2016}\\
$7334 \pm 10$   & $1.090$ & $7313 \pm 8$    & $1.049$ & $7334 \pm 10$   & $1.187$ & begin -- $7315$ & $1.049$ \\
$7392 \pm 33$   & $1.051$ & $7351 \pm 5$    & $1.054$ & $7393 \pm 32$   & $1.175$ & $7357 \pm 8$    & $1.030$ \\
$7553 \pm 16$   & $1.133$ & $7454 \pm 26$   & $1.022$ & $7551 \pm 21$   & $1.394$ & -               & -       \\
\multicolumn{8}{|c|}{observational season of 2016/2017}\\
begin -- $7743$ & $1.047$ & $7755 \pm 7$    & $1.082$ & begin -- $7739$ & $1.145$ & $7748 \pm 11$   & $1.027$ \\
$7773 \pm 10$   & $1.102$ & $7790 \pm 5$    & $1.080$ & $7773 \pm 10$   & $1.200$ & $7793 \pm 6$    & $1.026$ \\
$7831 \pm 33$   & $1.057$ & $7891 \pm 28$   & $1.032$ & $7831 \pm 31$   & $1.174$ & -               & -       \\
\hline
\end{tabular}
}
\label{tab_sched}
\end{center}
\end{table}

We can see series peaks in these plots, marking the position of optimal times.
In Table~\ref{tab_sched} we show more details concerning these optimal times for the
nearest three observation seasons.

\section{Conclusions}
A few interesting matters can be noticed in Fig.~\ref{fig_sched} and Table~\ref{tab_sched}:
\begin{enumerate}
\item The peaks of the efficiency functions are rather narrow, meaning that allocating
observation time randomly is not the best course of actions for HD~82943.

\item The task of refining the orbital inclination looks antagonistic to refining the
most other parameters. But nonetheless the relevant optimal time ranges tend
to stick together side-by-side.

\item We have good chances to refine the accuracy of the usual planetary parameters by
up to $30-40\%$ in the forthcoming observing seasons. But the orbital inclination, which is
only constrained thanks to the gravitational planet-planet perturbations, is an exception.
It looks unrealistic to drastically improve the accuracy of this inclination before 2020s,
when the orbital apsidal lines make a larger fraction of a secular revolution.

\item The refining of the parameters of the third planet seems rather difficult both in the
near and distant future. It seems that to reach this goal we should just patiently
accumulate more and more observations.
\end{enumerate}

\medskip
{\small The work was supported by the President of Russia grant for young scientists
(MK-733.2014.2), by the Russian Foundation for Basic Research (project 14-02-92615 KO\_a),
and by the programme of the Presidium of Russian Academy of Sciences
``Non-stationary phenomena in the objects of the Universe''.}


\begin{thebibliography}{}

\bibitem[Baluev (2008a)]{Baluev08a}
{Baluev, R.V.} 2008a,
\textit{MNRAS}, 389, 1375

\bibitem[Baluev (2008b)]{Baluev08b}
{Baluev, R.V.} 2008b,
\textit{Cel. Mech. \& Dyn. Astron.}, 102, 297

\bibitem[Baluev \& Beaug\'e (2014)]{BaluevBeauge14}
{Baluev, R.V., \& Beaug\'e, C.} 2014,
\textit{MNRAS}, 439, 673

\bibitem[Beaug\'e et al. (2003)]{Beauge03}
{Beaug\'e, C., Ferraz-Mello, S., \& Michtchenko, T.A.} 2003,
\textit{ApJ}, 593, 1124

\bibitem[Beaug\'e et al. (2008)]{Beauge08}
{Beaug\'e, C., Giuppone, C., Ferraz-Mello, S., \& Michtchenko T.A.} 2008,
\textit{MNRAS}, 385, 2151

\bibitem[Go\'zdziewski \& Konacki (2006)]{GozdKon06}
{Go\'zdziewski, K., \& Konacki, M.} 2006,
\textit{ApJ}, 647, 573

\bibitem[Mayor et al. (2004)]{Mayor04}
{Mayor, M. and Udry, S. and Naef, D. and Pepe, F. and Queloz, D. and Santos, N. C., \& Burnet, M.} 2004,
\textit{A\&A}, 415, 391

\bibitem[Tan et al. (2013)]{Tan13}
{Tan, X., Payne, M.J., Lee, M.H., Ford, E.B., Howard, A.W., Johnson, J.A., Marcy, G.W., \& Wright, J.T.} 2013,
\textit{ApJ}, 777, 101

\end{thebibliography}
\end{document}